\documentclass[twocolumn,english]{iopart}
\usepackage[T1]{fontenc}
\usepackage[latin9]{inputenc}
\usepackage{verbatim}
\usepackage{amstext}
\usepackage{amssymb}
\usepackage{graphicx}

\makeatletter
\usepackage{iopams}
\usepackage{setstack}

\makeatother

\usepackage{babel}
\begin{document}
\global\long\def\ket#1{\left|#1\right\rangle }

\global\long\def\bra#1{\left\langle #1\right|}

\global\long\def\braket#1#2{\left\langle #1\left|\vphantom{#1}#2\right.\right\rangle }

\global\long\def\ketbra#1#2{\left|#1\vphantom{#2}\right\rangle \left\langle \vphantom{#1}#2\right|}

\global\long\def\braOket#1#2#3{\left\langle #1\left|\vphantom{#1#3}#2\right|#3\right\rangle }

\global\long\def\mc#1{\mathcal{#1}}

\global\long\def\nrm#1{\left\Vert #1\right\Vert }

\title{On the equivalence of lossy evolution and POVM generalized quantum
measurements}

\title[On the equivalence of lossy evolution and POVM measurements]{}


\author{Raam Uzdin}

\address{Racah Institute of Physics, The Hebrew University, Jerusalem, Israel.}

\ead{raam@mail.huji.ac.il}
\begin{abstract}
Loss induced generalized measurements have been introduced years ago
as a mean to implement generalized quantum measurements (POVM). Here
the original idea is extended to a complete equivalence of lossy evolution
and a certain widely used class of POVM. This class includes POVM
used for unambiguous state discrimination and entanglement concentration.
One implication of this equivalence is that unambiguous state discrimination
schemes based on $\mc{PT}$ symmetric and non-Hermitian Hamiltonians
have the same performance as those of standard POVM. After discussing
several key points of this equivalence we illustrate our findings
in two elementary physical realizations. Finally, we discuss several
implications of this equivalence. 
\end{abstract}
\maketitle

\section{Introduction and motivation}

Positive Operator Valued Measure (POVM) \cite{Peres,Nielsen} refers
to a set of measurements within the standard theory of quantum mechanics
that generalizes the standard projective measurement. POVM's can be
viewed as standard projective quantum measurements in an augmented
Hilbert space (Neumark dilation \cite{Peres}). Using POVM's it is
possible to extract information that cannot be accessed by standard
measurements. Perhaps one of the most impressive examples of the power
of POVM is the unambiguous state discrimination (USD) problem (\cite{Jaeger Book,Barnett Book,Croke,Bergou SD rev,Chefles}
and references therein). USD has fundamental importance in quantum
information and quantum cryptography (\cite{Jaeger Book,Barnett Book}).
Consider a system that is prepared in one of the states $\ket{\alpha_{i}}$,
but not in a superposition of them. If the states are not orthogonal
($\braket{\alpha_{i}}{\alpha_{j\neq i}}\neq0$), there is no standard
projective measurement that can detect the state of the system without
an error that depends on the overlap of the states. In contrast, POVM,
can insure a zero error probability. As expected, this does not come
without a price. It turns out that there is an intrinsic nonzero chance
that the POVM will give an ``inconclusive result''. Namely, a result
that cannot be uniquely assigned to one of the input states (yet,
this is not considered an error). 

In \cite{Hutnner} Hutnner et al. suggested and demonstrated experimentally
the use of lossy evolution followed by projective measurements to
perform a state discrimination. In contrast to the dilated projectors
used in the Neumark scheme, their measurement is carried out in the
original Hilbert space. The implementation in \cite{Hutnner} is based
on embedding the lossy evolution in a larger Hilbert space where a
unitary evolution takes place. Bender et al. \cite{BB discr} suggested
a different scheme. Instead of embedding, a PT symmetric Hamiltonian
is used to generate a special non-unitary evolution that makes the
states orthogonal at the end of the evolution. Once the states are
orthogonal, a regular projective measurement can be used for discrimination.
This procedure also leads to zero error probability but so far it
has not been clarified how its performance (the inconclusive result
probability) compares to that of the best POVM. In a recent paper
\cite{NU resources}, the Hamiltonian resources needed for USD based
on $\mc{PT}$ symmetric/non-Hermitian Hamiltonian, have been quantified.
In this work, we focus on the success probability, and show that the
lossy (or $\mc{PT}$ symmetric/non-Hermitian generated) evolution
and the POVM state discrimination schemes are equivalent, and therefore
yield the same results. 

It should be mentioned that lossy evolution associated with non-Hermitian
Hamiltonians, is not a just a mathematical curiosity. It appears naturally
when the particle number in a subsystem of interest is not conserved.
In the past few years, non-Hermitian inspired non-unitary evolution
has been intensively studied experimentally and theoretically in the
context of $\mc{PT}$ symmetric Hamiltonians {[}11-19{]}. For other
studies of non-unitary evolution generated by more general Hamiltonians
see {[}20-28{]}.

We start by showing that given a lossy evolution there exists a USD
POVM that yields the same measurement results. This holds for any
lossy evolution regardless of state discrimination capabilities. Later
for completeness we repeat in greater detail and from a slightly different
point of view the converse direction that was studied in \cite{Hutnner}.
That is, given a USD-like POVM set, we show an explicit construction
of the equivalent lossy evolution operator. Several key implications
of this equivalence are shortly described at the end.

\section{Preliminaries}

\subsection{Projective measurements and optimal POVM}

In this section we describe POVM from the perspective of the pure
states USD problem. For a more general point of view we refer the
reader to \cite{Peres,Nielsen}. If a system is prepared in a normalized
state $\ket{\alpha_{i}}$ with probability $P_{i}$ the density matrix
is given by $\rho_{0}=\sum_{i=1}^{N}P_{i}\ketbra{\alpha_{i}}{\alpha_{i}}$.
In the USD problem the set $\{\ket{\alpha_{i}}\}_{i=1}^{N}$ is not
orthogonal. For generality we assume the complete USD problem where
in an $N$-level system, $N$ states should be discriminated. If there
are $L<N$ vectors of interest, other linearly-independent vectors
can be added artificially. Alternatively, the vectors of interest
can be unitarily rotated to a subspace of dimension L %
\footnote{This rotation can be accomplished by the following procedure. First
a Grahm-Schmidt orthogonalization on the input states is performed,
starting with the L vectors of interest. Then a unitary rotation is
used to rotate the new first L orthogonal unit vectors (that span
the original subspace L) to the computational basis $\{e_{i}\}_{i=1}^{L}$.
This transformation will also rotate the L vectors of interest to
the subspace $\{e_{i}\}_{i=1}^{L}$.%
} where once again the number of vectors is equal to the dimension
of the Hilbert space. 

Standard Von Neumann measurements are given by orthogonal projection
operators of the form $\Pi_{i}=\ketbra{\psi_{i}}{\psi_{i}}$ where
$\braket{\psi_{i}}{\psi_{j}}=\delta_{ij}$. From the orthogonality
of the states it follows that:
\begin{equation}
\Pi_{i}\Pi_{j}=\Pi_{i}\delta_{ij}.\label{eq: proj orthog}
\end{equation}
In addition $\sum_{i=1}^{N}\Pi_{i}=I$. The probability to find the
system in state the $\ket{\psi_{i}}$ is given by: $p_{i}=\mbox{tr}[\rho\Pi_{i}]$.
Since the system is prepared in one of the non-orthogonal states $\ket{\alpha_{i}}$,
there will be at least one state that will have a nonzero overlap
with more than one operator $\Pi_{i}$. Thus, when using $\Pi_{i}$,
an error in detecting the state of the system is inevitable. Yet,
it is possible to find a different set of rank-one operators that
will not overlap with more than one state. For this purpose we introduce
the bi-orthogonal set of states $\{\ket{\alpha_{i}^{\perp}}\}_{i=1}^{N}$
that satisfies:

\begin{equation}
\braket{\alpha_{i}^{\perp}}{\alpha_{j}}=\delta_{ij},\label{eq: bi orthog}
\end{equation}
The set $\{\ket{\alpha_{i}^{\perp}}\}$ can be obtained from $\{\ket{\alpha_{i}}\}$
in the following way. Let $A$ be a matrix whose columns are $\{\ket{\alpha_{i}}\}$.
Since $A^{-1}A=I$, the $\bra{\alpha_{i}^{\perp}}$ vectors are given
by the rows of $A^{-1}$. The transverse vectors are not orthogonal
to each other $\braket{\alpha_{i}^{\perp}}{\alpha_{j\neq i}^{\perp}}\neq0$.
Furthermore, while the vectors $\ket{\alpha_{i}}$ are normalized,
the vectors $\ket{\alpha_{i}^{\perp}}$ are not, and their amplitude
is determined by (\ref{eq: bi orthog}). In general, transverse vectors
are not uniquely defined. Yet, if $N$ transverse vectors are needed
in an $N$-dimensional Hilbert space, then the orthogonal vectors
are well defined (up to normalization) provided that the original
vector are linearly independent. 

From these transverse states we construct the rank one positive operators:
\begin{equation}
F_{i\le N}=\lambda_{i}\ketbra{\alpha_{i}^{\perp}}{\alpha_{i}^{\perp}},\label{eq: r1 POVM}
\end{equation}
where $0<\lambda_{i}$. We define another operator $F_{N+1}=I-\sum\lambda_{i}\ketbra{\alpha_{i}^{\perp}}{\alpha_{i}^{\perp}}$
so that together with $F_{i\le N}$ we get:
\begin{equation}
\sum_{i}^{N+1}F_{i}=I.\label{eq: completeness}
\end{equation}
The $\lambda_{i}$ can not be chosen completely arbitrarily since
$F_{N+1}$ must be a positive operator as well. A set of positive
operators that satisfies (\ref{eq: completeness}) is called a POVM.
In analogy to $\Pi_{i}$, the posterior probability to find the system
in the state $\ket{\alpha_{i}}$ (not $\ket{\alpha_{i}^{\perp}}$!)
by performing a POVM is: 
\begin{equation}
p_{i}=\mbox{tr}[\rho F_{i}].\label{eq: pk povm}
\end{equation}
The completeness relation (\ref{eq: completeness}) insures that $\sum p_{i}=1$
for any density matrix. Notice that relation (\ref{eq: proj orthog})
does not hold for POVM's, since the vectors $\ket{\alpha_{i}^{\perp}}$
are not orthogonal to each other. The $F_{i\le N}$ constructed here
are the POVM operators needed for the USD of states $\{\ket{\alpha_{i}}\}$.
The extra operator $F_{N+1}$ yields the probability of an inconclusive
result, $p_{N+1}$. A set of $\lambda_{i}$ that minimizes the inconclusive
results is called optimal \cite{Chefles}. In this work, the explicit
values of $\lambda_{i}$ are not needed. Notice that while $F_{i\le N}$
are rank one operators, while the rank of $F_{N+1}$ is typically
larger than one. 

In this work \textit{``USD} POVM'' refers to a POVM set $\{F_{i}\}_{i=1}^{N+1}$
over Hilbert space of dimension $N$, that has at least $N$ linearly
independent rank one operators. The purpose for which the POVM is
actually used may be different, but it still has the potential to
perform an $N$-state unambiguous discrimination.

\subsection{A lossy, single Kraus operator non-unitary evolution}

``Non-unitary'' evolution may refer to any evolution that is not
unitary. In this paper, however, we always refer to a specific type
of non-unitary evolution associated with losses (or in principle with
gain as well). Let $K(T)=K$ be an evolution operator so that a state
evolves from $t=0$ to $t=T$ according to: $\ket{\psi(T)}=K\ket{\psi(0)}$,
or in density matrix formalism: 
\begin{eqnarray}
\rho(T) & = & K\rho K^{\dagger}.\label{eq: single kraus}\\
K^{\dagger} & \neq & K^{-1}
\end{eqnarray}
A more general class of non-unitary evolution includes a sum over
different $K$'s . The most common scenario of this more general non-unitary
evolution, arises when decoherence terms are included in the Lindblad
equation \cite{Nielsen}. Yet, as shown here,\textit{ a single Kraus
operator (lossy evolution) is enough to establish a complete mapping
between USD POVM and non-unitary evolution}.

Unlike a general Kraus map, a ``Single Kraus operator'' evolution
(\ref{eq: single kraus}), can always be generated by the Schrödinger
equation with some non-Hermitian Hamiltonian. For example $\mc{PT}$
symmetric Hamiltonians generates evolution operators of the form (\ref{eq: single kraus}).
More generally, non-Hermitian Hamiltonians often appear in the study
of resonances and metastable systems \cite{Nim book}. Typical scenarios
include particle leakage from the system of interest (e.g. by tunneling)
or the presence of absorption in the medium. In optics, for example,
some photons are absorbed and converted into phonons. If only the
photons are of interest their effective description leads to a non-Hermitian
Hamiltonian (complex refractive index).

Although in optics the ``wavefunction'' can actually be amplified,
in quantum mechanics it is not so easily done. Therefore, we focus
here on passive systems, which cannot amplify the magnitude of the
state. The $K$ associated with such systems is characterized \cite{NU resources}
by $\nrm K_{sp}\le1$, where the spectral norm \cite{Horn}, $\nrm{\cdot}_{sp}$,
is equal to the largest singular value of $K$ (see matrix norm and
singular value decomposition in \cite{Horn}):

\begin{equation}
\nrm K_{sp}=\sqrt{max[eigenvalues(K^{\dagger}K)]}
\end{equation}
The passiveness condition becomes more apparent if an alternative
(yet equivalent) definition of the spectral norm is used. Let $\ket{\psi_{i}}$
be some initial state and $\ket{\psi_{f}}=K\ket{\psi_{i}}$ be a final
state. The spectral norm is given by:
\begin{equation}
\nrm K_{sp}=\underset{_{\ket{\psi_{i}}}}{\max}\sqrt{\frac{\braket{\psi_{f}}{\psi_{f}}}{\braket{\psi_{i}}{\psi_{i}}}}.
\end{equation}
 Namely, the spectral normal is the maximal amplitude amplification
$K$ can generate from all possible input states.

Another reason for looking at passive systems is that it makes the
performance comparison sensible. Otherwise, by controlling the signal
amplification, the detection probability can be effectively increased.
A lossy evolution can also be realized by embedding in a larger unitary
system \cite{Hutnner,Bergou embed 2000,Bergou embed 2006,Uwe neum brach,HNUE embedding}.
When only some part of Hilbert space is measured it appears as if
the evolution is non-unitary. In embedding schemes the condition $\nrm K_{sp}\le1$
is automatically satisfied.

\section{Equivalence of lossy evolution and USD POVM}

\subsection{The results of lossy coherent evolution can be reproduced by a POVM.}

Given some lossy evolution operator $K$, initial density matrix $\rho_{0}$,
and projective measurement operators $\Pi_{i}$, we want to show that
the probabilities $p_{i}=\mbox{tr}[\rho_{f}\Pi_{i}]=\mbox{tr}[K\rho_{0}K^{\dagger}\Pi_{i}]$
can be obtained by:
\begin{equation}
p_{i}=\mbox{tr}[\rho_{0}F_{i}],
\end{equation}
where $\{F_{i}\}_{i=1}^{N}$ is a part of a POVM set $\{F_{i\le N},F_{N+1}\}$
that satisfies the positivity and completeness (\ref{eq: completeness})
requirements. As explained before we assume $K$ corresponds to a
passive system (without gain) so $\nrm K_{sp}\le1$. This can always
be arranged by setting $K\to K/\Gamma$ where $\Gamma\ge\nrm K_{sp}$.
The probabilities at the end of the evolution are: 

\begin{eqnarray*}
p_{i}=\mbox{tr}[\rho_{f}\Pi_{i}] & = & \mbox{tr}[K\rho_{0}K^{\dagger}\Pi_{i}]=\mbox{tr}[\rho_{0}K^{\dagger}\Pi_{i}K]\triangleq\mbox{tr}[\rho_{0}F_{i}],
\end{eqnarray*}
Clearly the operators $F_{i}$ reproduce the same probabilities as
obtained by $K$. It remains to show that when complemented with another
operator $F_{n+1}$, the set $\{F_{n}\}_{n=1}^{N+1}$ constitutes
a legitimate POVM. We start by verifying that the $F_{i}$ operators
are positive. Using $\Pi_{i}^{2}=\Pi_{i}$ and $\Pi_{i}^{\dagger}=\Pi_{i}$:

\begin{equation}
F_{i}=K^{\dagger}\Pi_{i}K=(\Pi_{i}K)^{\dagger}\Pi_{i}K,\label{eq: POVM set N}
\end{equation}
it becomes clear that $F_{i}$ is positive since the RHS has the generic
form of a positive operator. Next, in order to satisfy (\ref{eq: completeness})
we define:
\begin{equation}
F{}_{N+1}=I-\sum_{i=1}^{N}F_{i}=I-K^{\dagger}K.\label{eq: POVM set N+1}
\end{equation}
Now (\ref{eq: completeness}) is trivially satisfied but it must be
verified that $F_{N+1}$ is a positive operator as well. In the diagonal
basis, $F_{N+1}$ is equal to $I-S^{2}$, where $S$ is a diagonal
positive matrix whose elements are the singular values of $K.$ Since
in passive systems the largest singular value ($\nrm K_{sp})$ must
be one or less, it follows that $F_{N+1}$ is positive (all of its
eigenvalues are non negative). This is a general feature; the passiveness
requirement of $K$ is equivalent to the POVM completeness requirement.
In summary, we conclude that:
\begin{equation}
\{F_{i\le N},F_{N+1}\}=\mbox{POVM}.\label{eq: POVM set part 2}
\end{equation}
The lossy evolution outcome is the same as that obtained from the
POVM set (\ref{eq: POVM set N}),(\ref{eq: POVM set N+1}). The immediate
consequence of the results above is that the performance of the $\mc{PT}$
symmetric discrimination scheme presented in \cite{BB discr} is the
same as the performance of the POVM scheme. Before concluding we point
out a feature of a POVM constructed from K. If the system is marginally
passive, $\nrm K_{sp}=1$, then $\text{rank}(F_{N+1})<N$. As a result
the Kraus operator $M_{n+1}$ defined by $F_{N+1}=M_{N+1}^{\dagger}M_{N+1}$
operator has a rank smaller than $N$. The new density matrix after
an inconclusive result is obtained, is $\rho_{?}=M_{N+1}\rho M_{N+1}^{\dagger}/\text{tr}(M_{N+1}\rho M_{N+1}^{\dagger})$.
Since the rank of $\rho_{?}$ is also smaller than $N$, we get that
the states in $\rho_{?}$ are no longer linearly independent%
\footnote{One may suspect that the rank reduction of $\rho_{?}$ with respect
to $\rho$ in the $\nrm K=1$ case is an indication the one of the
input states simply does not appear $\rho_{?}$ and as a result further
USD is possible. However in \cite{NUE disc singval} there is an argument
that explains why this is impossible.%
}. This means that the inconclusive density matrix cannot be used to
perform another \textit{Unambiguous} state discrimination \cite{lin indep Chefles}.

\subsection{USD POVM results can be reproduced by a lossy coherent evolution.}

In this subsection we show that any USD POVM in Hilbert space of size
$N$, is equivalent to a single lossy evolution operator (LEO) $K_{N\times N}$
followed by a \textit{projective} measurement. This should be contrasted
from the Neumark dilation scheme \cite{Peres} where POVM is interpreted/implemented
as projective measurements in a Hilbert space larger than $N$. To
avoid overlap with \cite{Hutnner} and to take a slightly more general
point of view we take a slightly different approach. Let $\left\{ F_{i}\right\} _{i=1}^{N+1}$
be a given USD POVM set where the first $N$ operators have rank one.
We forget for now about the state discrimination problem and consider
a bit more general problem. We want to replace $\left\{ F_{i}\right\} _{i=1}^{N+1}$
by an equivalent lossy evolution operator $K$ for \textit{any} density
matrix. That is, the density matrix to be measured may not be a statistical
mixture of the states $\left\{ F_{i}\right\} _{i=1}^{N+1}$ can discriminate.
Although the $\lambda_{i}$ and the $\ket{\alpha_{i}^{\perp}}$ normalization
can be calculated it is not explicitly needed. Therefore we can simply
write:
\begin{equation}
F_{i\le N}=\ketbra{\beta_{i}}{\beta_{i}},\label{eq: F beta}
\end{equation}
where the normalization of \textbf{$\ket{\beta_{i}}$ }is determined
by the given $F_{i}$. The extra operator satisfies $F_{N+1}=I-\sum_{i}^{N}F_{i}$. 

Given a density matrix, $\rho_{0}$, the probability to detect the
i-th result associated with the POVM operator $F_{i}$ is given by
$p_{i}=\mbox{tr}[\rho_{0}F_{i}]$. Our goal is to show that the same
probabilities can be obtained by a lossy evolution K:
\begin{equation}
p_{i}=\mbox{tr}[\rho_{f}\Pi_{i}]=\mbox{tr}[(K\rho_{0}K^{\dagger})\Pi_{i}],\label{eq: p_i SKO}
\end{equation}
where $\left\{ \Pi_{i}\right\} _{i=1}^{N}$ is a set of projective
measurement (i.e. satisfies (\ref{eq: proj orthog})) and $\rho_{f}$
is the final density operator generated by $K$. Next we find an explicit
expression for $K$ as a function of the chosen $\Pi_{i}$. From (\ref{eq: pk povm})
and (\ref{eq: p_i SKO}): 
\[
F_{i}=K^{\dagger}\Pi_{i}K.
\]
Direct substitution verifies that the $K$ that satisfies this relation
is:
\begin{equation}
K=\sum_{i=1}^{N}\frac{\pi_{i}F_{i}e^{\phi_{i}}}{\sqrt{tr(F_{i}\pi_{i})}},\label{eq: K pi F}
\end{equation}
where the $\phi_{i}$ are arbitrary phase degrees of freedom that
do not affect the measurement results or the measurement basis. Nonetheless
$\phi_{i}$ effects the eigenvalues of K and other properties of $K$.

\subsection{Properties of K}

In \cite{Hutnner}, it is mentioned that $K$ is diagonalizable operator
and that eigenvalues of $K$ have moduli smaller than one. It what
follows we clarify that the first statement is not necessary, and
that the second statement is not sufficient. The only limitations
on K are that $\nrm K_{sp}\le1$ and that $K$ is invertible. The
invertibility follows from the following argument. $K$ takes $N$
non-orthogonal linear independent vectors and transforms them into
$N$ orthogonal linearly independent vectors. Writing the vector in
column matrix G we have $G_{out}=KG_{in}$. Since $\text{det}(G_{in})\neq0,\text{det}(G_{out})\neq0$
if follows that $K$ must be invertible. \\
As an example of a legitimate non-diagonalizable lossy evolution operator
consider the following Jordan form evolution operator: $K_{J}=\left(\begin{array}{cc}
a & 1/2\\
0 & a
\end{array}\right)$. The passiveness condition is: $\left|a\right|\le1/\sqrt{2}$. The
states that can be discriminated are given by the columns of $(K_{J}^{-1})^{\dagger}$.
Furthermore, notice that the eigenvalues are just $a$ and for $1/\sqrt{2}<\left|a\right|\le1$
their modulus is smaller than one. Yet, in this regime the spectral
norm is larger than one and the evolution operator is not passive
anymore. In particular non-passive systems cannot be embedded in a
unitary evolution as suggested in \cite{Hutnner,Bergou embed 2006,Uwe neum brach,HNUE embedding}.
The necessary and sufficient equivalence condition $\nrm K_{sp}\le1$
does implies that the moduli of the eigenvalues of K are smaller than
one, but the converse is not true.

The lossy evolution operator (\ref{eq: K pi F}) can be written in
a more intuitive form. Using $\pi_{i}=\ketbra{\psi_{i}}{\psi_{i}}$
one can see that:
\begin{equation}
K=\sum_{i=1}^{N}a_{i}\ketbra{\psi_{i}}{\beta_{i}},
\end{equation}
where $a_{i}$ are some complex coefficients. Essentially, $K$ converts
the non-orthogonal vectors which are bi-orthonormal to $\ket{\beta}$
to the orthogonal states $\ket{\psi_{i}}$. Alternatively $K^{\dagger}$
takes the orthogonal states $\ket{\psi_{i}}$ to the non-orthogonal
states $\ket{\beta_{i}}$. This reflects the two complimentary points
of view on USD: one can think of $K$ as orthogonalization operator
that acts on the density matrix, or alternatively as an operator that
transform a standard projective measurement into a POVM. Further aspects
and properties of $K$ which are beyond the scope of this paper are
studied in \cite{NUE disc singval}.

\section{Illustrative physical examples }

In this section, we study two optical systems that demonstrate the
close kinship lossy evolution and POVM. For other USD implementations
in optics see \cite{Hutnner,Bergou embed 2000,Bergou embed 2006,opt real myers brandt 97,Brandt optics POVM,opt and sold disc}
and references therein. In the first example, we show how a lossy
evolution can implement a POVM without extending the Hilbert space,
while in the second example we examine an implementation of USD that
\textit{does} resort to Hilbert space dilation (embedding scheme). 

\begin{figure}
\includegraphics[width=12cm]{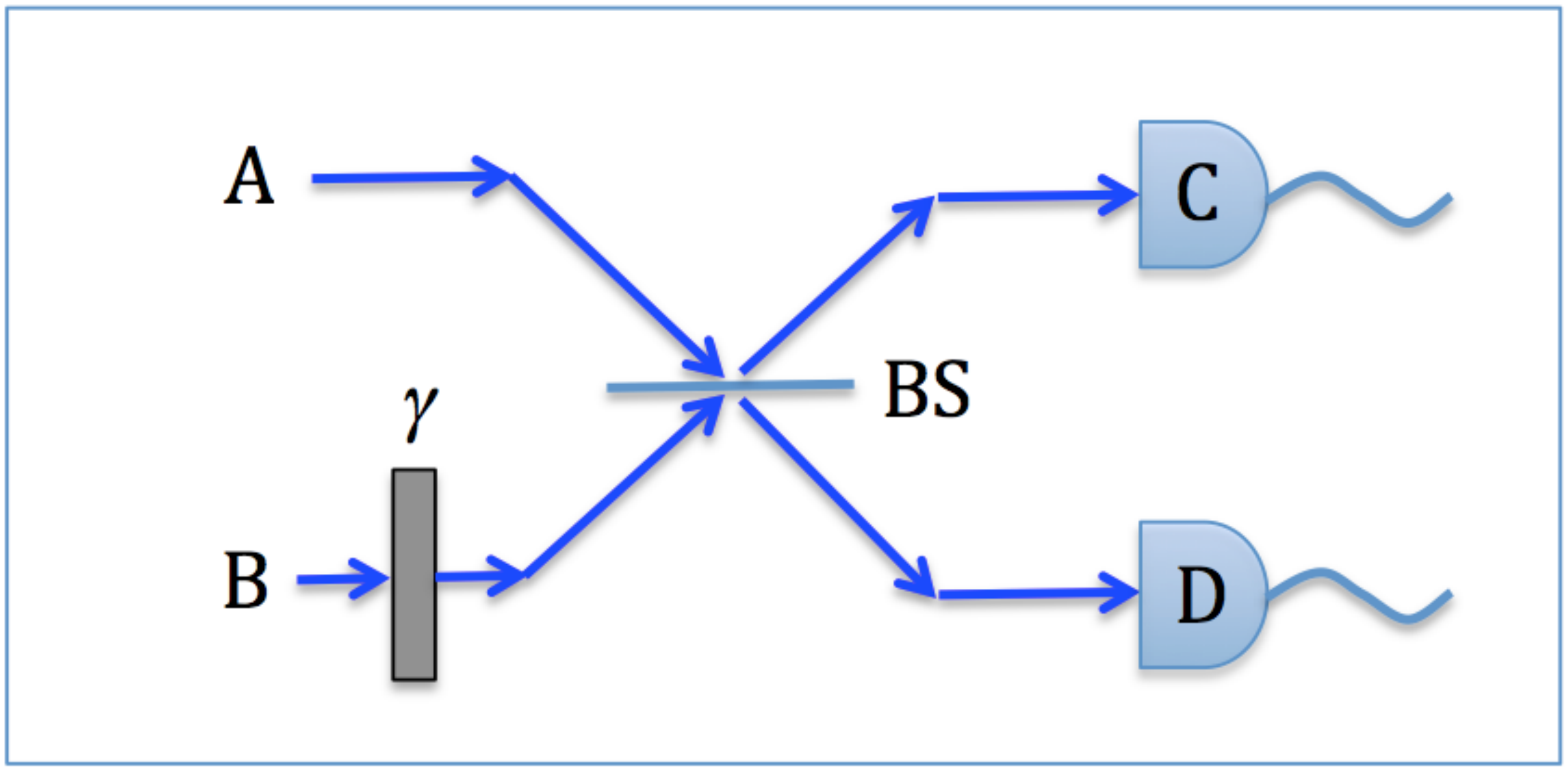}

\caption{a non-unitary optical implementation of a POVM measurement that performs
an unambiguous state discrimination (USD). A click at detectors C
or D will correctly indicate if the input is in state $\ket{\alpha_{1}}$
or $\ket{\alpha_{2}}$, even though $\left\langle \alpha_{2}|\alpha_{1}\right\rangle \neq0$.
This is possible only due to the presence of the attenuation plate
$\gamma$ that breaks unitarity. Since part of the light is absorbed,
in some cases there will be no click at the detectors. This is a manifestation
of the ``inconclusive result'' that appears in POVM based USD. }
\end{figure}
Figure 1 shows a very simple optical realization of a POVM using a
non-unitary element. The system consists of a 50-50 beam splitter
and an attenuator $\gamma$ that attenuates light by a factor $0<\gamma<1$.
The evolution operator is:
\begin{equation}
K=\frac{1}{\sqrt{2}}\left(\begin{array}{cc}
1 & \gamma\\
-1 & \gamma
\end{array}\right).
\end{equation}
We wish to find two non-orthogonal input vectors, $\ket{\alpha_{1,2}}$,
that at the output will populate exclusively either waveguide no.
1 (for $\ket{\alpha_{1}}$) or waveguide no. 2 (for $\ket{\alpha_{2}}$).
These vectors are given by the columns of $K^{\ensuremath{-1}}$ since
they need to satisfy: 
\begin{equation}
\left(\begin{array}{cc}
1 & 0\\
0 & 1
\end{array}\right)=K\{\ket{\alpha_{1}},\ket{\alpha_{2}}\}.\label{eq: inverse K}
\end{equation}
The LHS constitutes a choice of the measurement basis $\pi_{i}$.
For a different choice of $\pi_{i}=\ketbra{\psi_{i}}{\psi_{i}}$,
the identity matrix should be replaced by a matrix whose columns are
the $\ket{\psi_{i}}$ vectors. After normalization we get: $\ket{\alpha_{1,2}}=\frac{1}{\sqrt{1+\left|\gamma\right|^{2}}}(\pm\gamma,1)^{T}$
where '$T$' stands for transposition. Upon applying $K$ to these
vectors, the output is $\frac{\sqrt{2}\gamma}{\sqrt{1+\left|\gamma\right|^{2}}}(1,0)^{T}$
for $\ket{\alpha_{1}}$, and $\frac{\sqrt{2}\gamma}{\sqrt{1+\left|\gamma\right|^{2}}}(0,1)^{T}$
for $\ket{\alpha_{2}}$. Since, at the output the vectors are orthogonal,
they can easily be discriminated by detectors C and D. Notice that
as the input vectors become almost identical ($\gamma\to0$) the detection
probability goes to zero, \textit{since the output is proportional
to $\gamma$ for these specific input vectors}. Though understandable,
we find it beautiful that loosing part of the input signal gives access
to information that lies outside the reach of unitary evolution. 

\begin{figure}
\includegraphics[width=12cm]{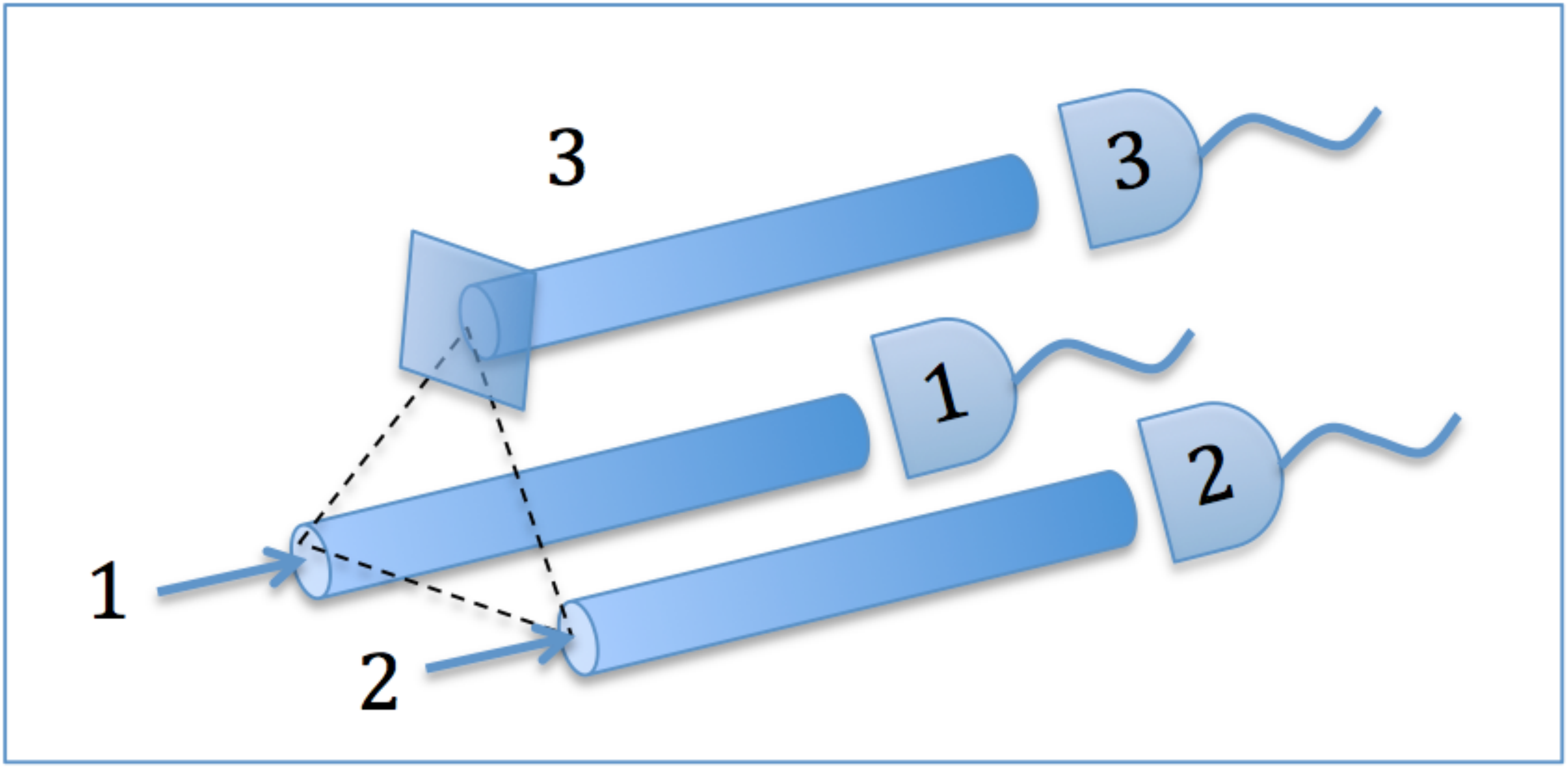}

\caption{In this system, the non-unitarity is achieved by coupling the light
to an auxiliary waveguide (no. 3) that is not initially populated.
The effective evolution of light in the subsystem of waveguides 1
and 2 is non-unitary, and USD becomes possible. A click at detectors
1 or 2 means a successful discrimination while a click at the third
detector means an inconclusive result. }
\end{figure}
Figure 2 shows another optical system that consists of three parallel
waveguides equally spaced from each other. If the waveguides are not
too close to each other, this system is well described by the tight
binding Hamiltonian:
\begin{equation}
H=a\left(\begin{array}{ccc}
0 & 1 & 1\\
1 & 0 & 1\\
1 & 1 & 0
\end{array}\right),
\end{equation}
where the components of the state vector are the peak amplitudes in
waveguides 1,2 and 3. Without loss of generality, we set $a=1$ (it
just rescales the propagation coordinate). Quantum mechanically, this
Hamiltonian can describe three potential wells arranged in an equilateral
triangular where only the ground state interaction is dominant (i.e.
there is a large energy gap to the next level). We denote by '1' and
'2' the input ports and the output ports of interest. Waveguide no.
3 will be used as an auxiliary waveguide that is initially not populated.
The unitary evolution of the three waveguides is given by $U=exp(-iHz)$,
where $z$, the propagation coordinate, plays the role of time. By
applying $U$ to $(1,0,0)$ and to $(0,1,0)$, we can construct a
reduced two-waveguide evolution operator $K=K_{2\times2}(z)$. This
$K$ will give the correct output of waveguides 1 and 2 for any input
state that does not initially populate the third waveguide. In general,
$K$ is not unitary, since part of the optical power goes to waveguide
no. 3. According to (\ref{eq: inverse K}), the input states the system
is able to discriminate, can be obtained from $K^{-1}$. To see the
evolution in the whole three-waveguide system, we apply $U$ (instead
of $K$) on these two input states. One can show that the output is
of the form:

\begin{eqnarray}
U(\ket{u_{1}}^{T},0)^{T} & = & (\beta,0,\sqrt{1-\mbox{\ensuremath{\left|\beta\right|}}^{2}}),\label{eq: out vec 1}\\
U(\ket{u_{2}}^{T},0)^{T} & = & (0,\beta,\sqrt{1-\mbox{\ensuremath{\left|\beta\right|}}^{2}}).\label{eq: out vec 2}
\end{eqnarray}
 The factor $\left|\beta\right|\le1$ becomes smaller as $\left|\braket{\alpha_{1}}{\alpha_{2}}\right|$
becomes larger (i.e. when the input vectors are more similar to each
other). To complete the USD scheme, a photon detector is placed at
the output of each port. If there is a hit at no. 1 (no. 2) we infer
the system was in state $\ket{\alpha_{1}}$ ($\ket{\alpha_{2}}$).
If there is a hit at detector no. 3, we cannot tell what was the state
of the system (follows from the form of (\ref{eq: out vec 1}) and
(\ref{eq: out vec 2})). This is exactly the POVM inconclusive result.
We conclude that this simple apparatus successfully implements USD
for ports 1 and 2. Inspecting the output vectors (\ref{eq: out vec 1})
and (\ref{eq: out vec 2}), we see that, as expected, they remained
non-orthogonal when all three components are considered, since $U$
is unitary. Yet, when only the subspace 1 and 2 is considered, the
two vectors look orthogonal at the output. 

Notice that there is a relation to the first example. If the attenuator
is replaced by a beam splitter with transmittance $\gamma$, then
the inconclusive result can be detected by monitoring the reflected
photon. By adding this extra port, the system is now described by
a larger Hilbert space just like in the second example.

\section{Concluding remarks}

The implications of the results presented here extend beyond the formal
equivalence of two different approaches to unambiguous state discrimination.
In \cite{HNUE embedding} we discuss the resources needed for embedding
a lossy evolution in a larger Hilbert space where a unitary (zero-loss)
evolution takes place. Together with the findings presented here we
obtain a non-trivial relation between energy and generalized measurements.
We find what are the minimal Hamiltonian resources needed to embed
a USD POVM in a unitary evolution.

A second implication concerns the general theory of multiple quantum
state discrimination. The representation of a USD POVM set of operators
by a single lossy evolution operator reveals new features of multiple
state discrimination that are very difficult to deduce directly from
the original POVM set of operators \cite{NUE disc singval}.

\ack{}{The author thanks Omri Gat for useful comments. \protect \\
}

\end{document}